\input{aipcheck}

\documentclass[
,draft            
,numberedheadings 
  ]
  {aipproc}

\layoutstyle{8x11single}

\begin{document}

\title{Stable phantom energy traversable wormhole models}

\classification{04.20.Gz, 04.20.Jb, 98.80.Es}

\keywords {wormholes, phantom energy, junction formalism}

\author{Francisco S. N. Lobo}{
  address={Centro de Astronomia
e Astrof\'{\i}sica da Universidade de Lisboa, \\
Campo Grande, Ed. C8 1749-016 Lisboa, Portugal} }

\begin{abstract}

A possible candidate for the present accelerated expansion of the
Universe is ``phantom energy'', which possesses an equation of
state of the form $\omega\equiv p/\rho<-1$, consequently violating
the null energy condition. As this is the fundamental ingredient
to sustain traversable wormholes, this cosmic fluid presents us
with a natural scenario for the existence of these exotic
geometries. In this context, we shall construct phantom wormhole
geometries by matching an interior wormhole solution, governed by
the phantom energy equation of state, to an exterior vacuum at a
junction interface. Several physical properties and
characteristics of these solutions are further investigated. The
dynamical stability of the transition layer of these phantom
wormholes to linearized spherically symmetric radial perturbations
about static equilibrium solutions is also explored. It is found
that the respective stable equilibrium configurations may be
increased by strategically varying the wormhole throat radius.

\end{abstract}

\maketitle


\section{Introduction}

One of the most challenging current problems in cosmology is
explaining the accelerated phase of expansion of the Universe
\cite{expansion}. Several candidates, responsible for this cosmic
expansion, have been proposed in the literature, namely, dark
energy models, generalizations of the Chaplygin gas, modified
gravity and scalar-tensor theories, tachyon scalar fields and
braneworld models, such as the Dvali-Gabadadze-Porrati (DGP)
model, amongst others. The dark energy models are parametrized by
an equation of state given by $\omega =p/\rho$, where $p$ is the
spatially homogeneous pressure and $\rho$ is the dark energy
density. For the cosmic expansion, a value of $\omega<-1/3$ is
required, as dictated by the Friedman equation $\ddot{a}/a=-4\pi
(p+\rho/3)$, so that $\ddot{a}>0$.
A specific exotic form of dark energy, denoted phantom energy, has
also been proposed, possessing the peculiar property of
$\omega<-1$ \cite{Weinberg}. This parameter range is not excluded
by observation, and possesses peculiar properties, such as the
violation of the null energy condition and an infinitely
increasing energy density, resulting in a Big Rip, at which point
the Universe blows up in a finite time~\cite{Weinberg}. However,
recent fits to supernovae, CMB and weak gravitational lensing data
indicate that an evolving equation of state $\omega$ crossing the
phantom divide $-1$, is mildly favored \cite{phantom-divide}.

As the phantom energy equation of state violates the null energy
condition, $p+\rho<0$, which is the fundamental ingredient to
sustain traversable wormhole~\cite{Morris,Visser}, one now has at
hand a possible source for these exotic spacetimes. However, a
subtlety needs to be pointed out, as emphasized in Refs.
\cite{Sushkov,Lobo-phantom}. The notion of phantom energy is that
of a homogeneously distributed fluid. When extended to
inhomogeneous spherically symmetric spacetimes, the pressure
appearing in the equation of state is now a radial pressure, and
the transverse pressure is then determined via the field
equations. Wormhole geometries were found in Ref. \cite{Sushkov},
by considering specific choices for the distribution of the energy
density, and in Ref. \cite{Lobo-phantom}, a complementary approach
was traced out, by imposing appropriate choices for the form
function and/or the redshift function, and the stress-energy
tensor components were consequently determined. This latter
approach shall be traced out in this paper.

It is also of a fundamental importance to investigate the
stability of these phantom wormhole geometries. We shall model
these spacetimes by matching an interior traversable wormhole
geometry with an exterior Schwarzschild vacuum solution at a
junction interface \cite{Lobo-CQG}. In this work, we analyze the
stability of these phantom wormholes to linearized perturbations
around static solutions. Work along these lines was done by
considering thin-shell wormholes, using the cut-and-paste
technique \cite{Poisson}. The advantage of this analysis resides
in using a parametrization of the stability of equilibrium, so
that there is no need to specify a surface equation of state. The
stability analysis of these thin-shell wormholes to linearized
spherically symmetric perturbations about static equilibrium
solutions was carried out by assuming that the shells remain
transparent under perturbation \cite{Ishak}. This amounts to
considering specific spacetimes that do not contribute with the
momentum flux term in the conservation identity, which provides
the conservation law for the surface stress-energy tensor. The
inclusion of this term, corresponding to the discontinuity of the
momentum impinging on the shell, severely complicates the
analysis. However, we shall follow the approach of Ishak and Lake
\cite{Ishak}, with the respective inclusion of the momentum flux
term and find stability equilibrium configurations of specific
phantom wormhole geometries.

This paper is organized in the following manner. In Section
\ref{Sec:phantomWH}, we outline the general properties of phantom
energy traversable wormhole, and in Section \ref{Sec:specificWH},
we present specific solutions of these phantom wormholes. We also
show that using these specific constructions, and taking into
account the ``volume integral quantifier'', one may theoretically
construct these spacetimes with infinitesimal amounts of ANEC
violating phantom energy. In Section \ref{Sec:stability}, we
outline a general linearized stability analysis procedure, and
then apply this analysis to phantom wormhole geometries and
determine their respective stability regions. Finally in Section
\ref{Sec:conclusion}, we conclude.

\section{Phantom energy traversable wormholes}\label{Sec:phantomWH}


The spacetime metric representing a spherically symmetric and
static wormhole is given by
\begin{equation}
ds^2=-e ^{2\Phi(r)}\,dt^2+\frac{dr^2}{1- b(r)/r}+r^2 \,(d\theta
^2+\sin ^2{\theta} \, d\phi ^2) \label{metricwormhole}\,,
\end{equation}
where $\Phi(r)$ and $b(r)$ are arbitrary functions of the radial
coordinate, $r$. $\Phi(r)$ is denoted as the redshift function,
for it is related to the gravitational redshift; $b(r)$ is called
the form function \cite{Morris}. The radial coordinate has a range
that increases from a minimum value at $r_0$, corresponding to the
wormhole throat, to $a$, where the interior spacetime will be
joined to an exterior vacuum solution.

Using the Einstein field equation, $G_{\hat{\mu}\hat{\nu}}=8\pi
\,T_{\hat{\mu}\hat{\nu}}$, in an orthonormal reference frame,
(with $c=G=1$) we obtain the following stress-energy scenario
\begin{eqnarray}
b'&=&8\pi r^2 \rho  \label{rhoWH}\,,\\
\Phi'&=&\frac{b+8\pi r^3 p_r}{2r^2(1-b/r)}  \label{prWH}\,,\\
p_r'&=&\frac{2}{r}\,(p_t-p_r)-(\rho +p_r)\,\Phi ' \label{ptWH}\,,
\end{eqnarray}
where the prime denotes a derivative with respect to the radial
coordinate, $r$. $\rho(r)$ is the energy density, $p_r(r)$ is the
radial pressure, and $p_t(r)$ is the lateral pressure measured in
the orthogonal direction to the radial direction. Equation
(\ref{ptWH}) may be obtained using the conservation of the
stress-energy tensor, $T^{\mu\nu}_{\;\;\;\;;\nu}=0$. At the throat
we have the flaring out condition given by $(b-b'r)/b^2>0$
\cite{Morris,Visser}, which may be deduced from the mathematics of
embedding. From this we verify that at the throat $b(r_0)=r=r_0$,
the condition $b'(r_0)<1$ is imposed to have wormhole solutions.
For the wormhole to be traversable, one must demand that there are
no horizons present, which are identified as the surfaces with
$e^{2\Phi}\rightarrow 0$, so that $\Phi(r)$ must be finite
everywhere. Note that the condition $1-b/r>0$ is also imposed.

A fundamental ingredient of traversable wormholes and phantom
energy is the violation of the null energy condition (NEC). Matter
that violates the NEC is denoted as {\it exotic matter}. Note that
the notion of phantom energy is that of a homogeneously
distributed fluid; however, when extended to inhomogeneous
spherically symmetric spacetimes, the pressure appearing in the
equation of state is now a radial pressure, and the transverse
pressure is then determined from Eq. (\ref{ptWH}). Using the
equation of state representing phantom energy, $p_r=\omega \rho$
with $\omega<-1$, and taking into account Eqs.
(\ref{rhoWH})-(\ref{prWH}), we have the following condition
\begin{equation}
\Phi'(r)=\frac{b+\omega rb'}{2r^2\,\left(1-b/r \right)} \,.
            \label{EOScondition}
\end{equation}
We now have four equations, namely the field equations, i.e., Eqs.
(\ref{rhoWH})-(\ref{ptWH}), and Eq. (\ref{EOScondition}), with
five unknown functions of $r$, i.e., $\rho(r)$, $p_r(r)$,
$p_t(r)$, $b(r)$ and $\Phi(r)$. To construct solutions with the
properties and characteristics of wormholes, we consider
restricted choices for $b(r)$ and/or $\Phi(r)$. In cosmology the
energy density related to the phantom energy is considered
positive, $\rho>0$, so we shall maintain this condition. This
implies that only form functions of the type $b'(r)>0$ are
allowed. We can construct asymptotically flat spacetimes, in which
$b(r)/r\rightarrow 0$ and $\Phi\rightarrow 0$ as $r\rightarrow
\infty$. However, one may also construct solutions with a cut-off
of the stress-energy, by matching the interior solution to an
exterior vacuum spacetime, at a junction interface.

\section{Specific phantom wormhole models}\label{Sec:specificWH}

\subsection{Asymptotically flat spacetimes}

To construct an asymptotically flat wormhole solution
\cite{Lobo-phantom}, consider $\Phi(r)=\Phi_0={\rm const}$. Thus,
from Eq. (\ref{EOScondition}) one obtains
\begin{equation}
b(r)=r_0(r/r_0)^{-1/\omega}  \,,
   \label{flat-form}
\end{equation}
so that $b(r)/r=(r_0/r)^{(1+\omega)/\omega}\;\rightarrow 0$ for
$r\rightarrow \infty$. We also verify that
$b'(r)=-(1/\omega)(r/r_0)^{-(1+\omega)/\omega}$, so that at the
throat the condition $b'(r_0)=1/|\omega|<1$ is satisfied.

The stress-energy tensor components are given by
\begin{eqnarray}
p_r(r)=\omega \rho(r) =-\frac{1}{8\pi
r_0^2}\left(\frac{r_0}{r}\right)^{3+\frac{1}{\omega}}  \,,
 \qquad {\rm and}  \qquad
p_t(r)=\frac{1}{16\pi r_0^2}\left(\frac{1+\omega}{\omega}\right)
\left(\frac{r_0}{r}\right)^{3+\frac{1}{\omega}} \,.
\end{eqnarray}
Thus, determining the parameter $\omega$ from observational
cosmology, assuming the existence of phantom energy, one may
theoretically construct traversable phantom wormholes by
considering the above-mentioned form function and a constant
redshift function.

Using the ``volume integral quantifier'', which provides
information about the ``total amount'' of averaged null energy
condition (ANEC) violating matter in the spacetime (see Ref.
\cite{VKD1} for details), given by
$I_V=\int\left[\rho(r)+p_r(r)\right]dV$, with a cut-off of the
stress-energy at $a$, we have
\begin{eqnarray}
I_V=\int_{r_0}^a
\left(r-b\right)\left[\ln\left(\frac{e^{2\Phi}}{1-b/r}\right)\right]'\;dr
\,.
    \label{vol-int}
\end{eqnarray}

Considering the specific choices for the form function and
redshift function for the traversable wormhole given above, with
$\omega=-2$ for simplicity, the volume integral assumes the value
$I_V= r_0 (1-\sqrt{a/r_0})$. Taking the limit $a \rightarrow r_0$,
one verifies that $I_V=\int (\rho + p_r)\; dV \rightarrow 0$.
Thus, as in the examples presented in \cite{VKD1}, one may
construct a traversable wormhole with arbitrarily small quantities
of ANEC violating phantom energy. Although this result is not
unexpected it is certainly a fascinating prospect that an advanced
civilization may probably construct and sustain a wormhole with
vanishing amounts of the material that comprises of approximately
70\% of the constitution of the Universe.

\subsection{Isotropic pressure}

Consider now isotropic pressures, $p_r=p_t=p$. By taking into
account the form function, considered above, given by
$b(r)=r_0\,(r/r_0)^{-1/\omega}$, and using Eq. (\ref{ptWH}) and
Eq. (\ref{rhoWH}), one finds that the redshift function is given
by
\begin{equation}
\Phi(r)=\left(\frac{3\omega+1}{1+\omega}\right)\;\ln
\left(\frac{r}{r_0}\right) \,.
   \label{isotropic-redshift}
\end{equation}
(See Ref. \cite{Lobo-phantom} for details). The stress-energy
tensor components are provided by
\begin{eqnarray}
p(r)&=&\omega \rho(r) =-\frac{1}{8\pi
r_0^2}\left(\frac{r_0}{r}\right)^{3+\frac{1}{\omega}} \,.
\end{eqnarray}

Note that the spacetime is not asymptotically flat. However, one
may match the interior wormhole solution to an exterior vacuum
spacetime at a finite junction surface. Using the ``volume
integral quantifier'', Eq. (\ref{vol-int}), with a cut-off of the
stress-energy at $a$, and taking into account the choices for the
form function and redshift function considered above, with
$\omega=-2$, the volume integral takes the following value
\begin{eqnarray}
I_V= a \left(1-\sqrt{r_0/a} \; \right)\;\ln
\left[\frac{(a/r_0)^{10}}{1-\sqrt{r_0/a}} \; \right]
+\left(10a+11r_0-21r_0\sqrt{a/r_0} \; \right)+a
\left(\sqrt{r_0/a}-1 \; \right)\;\ln
\left[\frac{(a/r_0)^{21/2}}{\sqrt{a/r_0}-1} \; \right] \,.
\end{eqnarray}
Once again taking the limit $a \rightarrow r_0$, one verifies that
$I_V \rightarrow 0$, and as before one may construct a traversable
wormhole with arbitrarily small quantities of ANEC violating
phantom energy.

\section{Stability analysis}\label{Sec:stability}

\subsection{Junction conditions}

We shall model specific phantom wormholes by matching an interior
traversal wormhole geometry, satisfying the equation of state
$p_r=\omega \rho$ with $\omega<-1$, with an exterior Schwarzschild
solution at a junction interface $\Sigma$, situated outside the
event horizon, $a>r_b=2M$.

Using the Darmois-Israel formalism \cite{Israel}, the surface
stress-energy tensor, $S^i{}_j$, at the junction interface
$\Sigma$, provide us with the following expressions for the
surface stresses
\begin{eqnarray}
\sigma&=&-\frac{1}{4\pi a} \left(\sqrt{1-\frac{2M}{a}+\dot{a}^2}-
\sqrt{1-\frac{b(a)}{a}+\dot{a}^2} \, \right)
    \label{surfenergy}   ,\\
{\cal P}&=&\frac{1}{8\pi a} \Bigg[\frac{1-\frac{M}{a}
+\dot{a}^2+a\ddot{a}}{\sqrt{1-\frac{2M}{a}+\dot{a}^2}}
   -\frac{(1+a\Phi') \left(1-\frac{b}{a}+\dot{a}^2
\right)+a\ddot{a}-\frac{\dot{a}^2(b-b'a)}{2(a-b)}}{\sqrt{1-\frac{b(a)}{a}+\dot{a}^2}}
\, \Bigg]         \,,
    \label{surfpressure}
\end{eqnarray}
where $\sigma$ and ${\cal P}$ are the surface energy density and
the tangential surface pressure, respectively (see Refs.
(\cite{stable-phanWH,stable-darkstar,Lobo-Craw}) for details).

We shall use the conservation identity, given by
$S^{i}_{j|i}=\left[T_{\mu \nu}e^{\mu}_{(j)}n^{\nu}\right]^+_-$,
where $n^{\mu}$ is the unit normal $4-$vector to $\Sigma$, and
$e^{\mu}_{(i)}$ are the components of the holonomic basis vectors
tangent to $\Sigma$. The momentum flux term in the right hand side
corresponds to the net discontinuity in the momentum which
impinges on the shell.
Using $S^{i}_{\tau|i}=-\left[\dot{\sigma}+2\dot{a}(\sigma +{\cal
P} )/a \right]$, the conservation identity provides us with
\begin{equation}
\sigma'=-\frac{2}{a}\,(\sigma+{\cal P})+\Xi  \qquad {\rm with}
\qquad      \Xi=-\frac{1}{4\pi a^2}
\left[\frac{b'a-b}{2a\left(1-\frac{b}{a} \right)}+a\Phi' \right]
\sqrt{1-\frac{b}{a}+\dot{a}^2} \,.
  \label{consequation2}
\end{equation}
Equation (\ref{consequation2}) evaluated at a static solution
$a_0$, shall play a fundamental role in determining the stability
regions.

Equation (\ref{surfenergy}) may be rearranged to provide the thin
shell's equation of motion, i.e., $\dot{a}^2 + V(a)=0$, with the
potential given by
\begin{equation}
V(a)=F(a)-\left(\frac{m_s}{2a}\right)^2-\left(\frac{aG}{m_s}\right)^2
\,,
\end{equation}
where $m_s=4\pi \sigma a^2$ is the surface mass of the thin shell.
For computational purposes and notational convenience, we define
the following factors
\begin{eqnarray}
F(a)=1-\frac{b(a)/2+M}{a} \qquad {\rm and} \qquad
G(a)=\frac{M-b(a)/2}{a} \,,
\end{eqnarray}

Linearizing around a stable solution situated at $a_0$, we
consider a Taylor expansion of $V(a)$ around $a_0$ to second
order, given by
\begin{eqnarray}
V(a)=V(a_0)+V'(a_0)(a-a_0)
+\frac{1}{2}V''(a_0)(a-a_0)^2+O[(a-a_0)^3] \,.
\label{linear-potential}
\end{eqnarray}
Evaluated at the static solution, at $a=a_0$, we verify that
$V(a_0)=0$ and $V'(a_0)=0$. From the condition $V'(a_0)=0$, one
extracts the following useful equilibrium relationship
\begin{eqnarray}
\Gamma\equiv\left(\frac{m_s}{2a_0}\right)'
=\left(\frac{a_0}{m_s}\right)\left[
F'-2\left(\frac{a_0G}{m_s}\right)\left(\frac{a_0G}{m_s}\right)'\right]
  \,,
\end{eqnarray}
which will be used in determining the master equation, responsible
for dictating the stable equilibrium configurations.

The solution is stable if and only if $V(a)$ has a local minimum
at $a_0$ and $V''(a_0)>0$ is verified. The latter condition takes
the form
\begin{eqnarray}
\left(\frac{m_s}{2a}\right)\left(\frac{m_s}{2a}\right)''<\Psi
-\Gamma^2 \,, \qquad {\rm with} \qquad
\Psi=\frac{F''}{2}-\left[\left(\frac{aG}{m_s}\right)'\right]^2
-\left(\frac{aG}{m_s}\right)\left(\frac{aG}{m_s}\right)''  \,.
     \label{masterequation}
\end{eqnarray}
Using $m_s=4\pi a^2 \sigma$, and taking into account the radial
derivative of $\sigma'$, Eq. (\ref{consequation2}) can be
rearranged to provide the following relationship
\begin{equation}
\left(\frac{m_s}{2a}\right)''= \Upsilon -4\pi \sigma'\eta \,,
\qquad {\rm with} \qquad    \Upsilon \equiv
\frac{4\pi}{a}\,(\sigma+{\cal P})+2\pi a \, \Xi ' \,,
     \label{cons-equation2}
\end{equation}
and the parameter $\eta$ is defined as $\eta={\cal P}'/\sigma'$.

Equation (\ref{cons-equation2}) will play a fundamental role in
determining the stability regions of the respective solutions.
Note that the parameter $\sqrt{\eta}$ is normally interpreted as
the speed of sound, so that one would expect that $0<\eta \leq 1$,
based on the requirement that the speed of sound should not exceed
the speed of light. However, in the presence of exotic matter this
cannot naively de done so. Therefore, in this work the above range
will be relaxed. We refer the reader to Ref. \cite{Poisson} for an
extensive discussion on the respective physical interpretation of
$\eta$ in the presence of exotic matter.

We shall use $\eta$ as a parametrization of the stable
equilibrium, so that there is no need to specify a surface
equation of state. Thus, substituting Eq. (\ref{cons-equation2})
into Eq. (\ref{masterequation}), one deduces the master equation
given by
\begin{equation}
\sigma' \,m_s \,\eta_0 > \Theta\,,  \qquad {\rm with} \qquad
\Theta \equiv \frac{a_0}{2\pi} \left(\Gamma^2-\Psi \right)
+\frac{1}{4\pi}\,m_s\,\Upsilon   \,,
\end{equation}
and $\eta_0=\eta(a_0)$. The master equation dictate stable
equilibrium regions, given by the following inequalities
\begin{eqnarray}
\eta_0 &>& \overline{\Theta}, \qquad {\rm if} \qquad \sigma'
\,m_s>0\,,      \label{stability1}
       \\
\eta_0 &<& \overline{\Theta}, \qquad {\rm if} \qquad \sigma'
\,m_s<0\,,       \label{stability2}
\end{eqnarray}
with the definition $\overline{\Theta}\equiv \Theta/(\sigma'
\,m_s)$.

We shall now consider the phantom wormhole geometries found in
Secion \ref{Sec:specificWH}, and consequently determine the
stability regions dictated by the inequalities
(\ref{stability1})-(\ref{stability2}). In the specific cases that
follow, the explicit form of $\Upsilon$, $\Psi$, $\Theta$ and
$\overline{\Theta}$ are extremely messy, so that as in
\cite{Ishak}, we find it more instructive to show the stability
regions graphically.

\subsection{Stability regions}

\subsubsection{Asymptotically flat spacetimes}

Consider the specific choices for the redshift and form functions,
for an asymptotically flat spacetime, given by
$\Phi(r)=\Phi_0={\rm const}$  and the form function given by Eq.
(\ref{flat-form}), $b(r)=r_0(r/r_0)^{-1/\omega}$.
To determine the stability regions of this solution, we shall
separate the cases of $b(a_0)<2M$ and $b(a_0)>2M$. From Eq.
(\ref{surfenergy}) and the definition of $m_s=4\pi a_0^2 \sigma$,
this corresponds to $m_s >0$ and $m_s <0$, respectively. Here, we
shall relax the condition that the surface energy density be
positive, as in considering traversable wormhole geometries, one
is already dealing with exotic matter. Note that for $\sigma<0$,
the weak energy condition is readily violated.

For $b(a_0)<2M$, i.e., for a positive surface energy density, we
need to impose the condition $r_0<2M$, so that the junction radius
lies outside the event horizon, $a_0>2M$. Thus, the junction
radius lies in the following range
\begin{equation}
2M< a_0 <2M \left(2M/r_0\right)^{-(1+\omega)}  \,.
     \label{flat-range}
\end{equation}
For a fixed value of $\omega$, we verify that as $r_0 \rightarrow
0$, then $a_0 \rightarrow \infty$. The range decreases, i.e., $a_0
\rightarrow 2M$, as $r_0 \rightarrow 2M$. Note that by fixing
$r_0$ and decreasing $\omega$, the range of $a_0$ is also
significantly increased.
\begin{figure}[t]
\centering
  \includegraphics[width=2.3in]{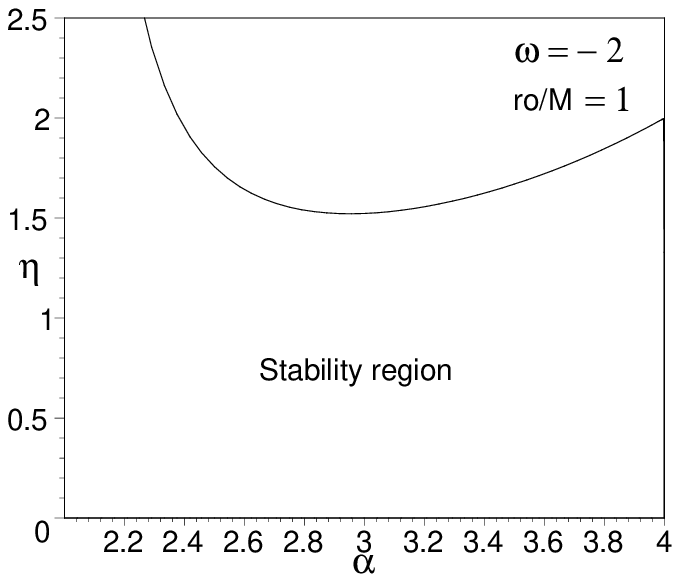}
  \hspace{0.4in}
  \includegraphics[width=2.3in]{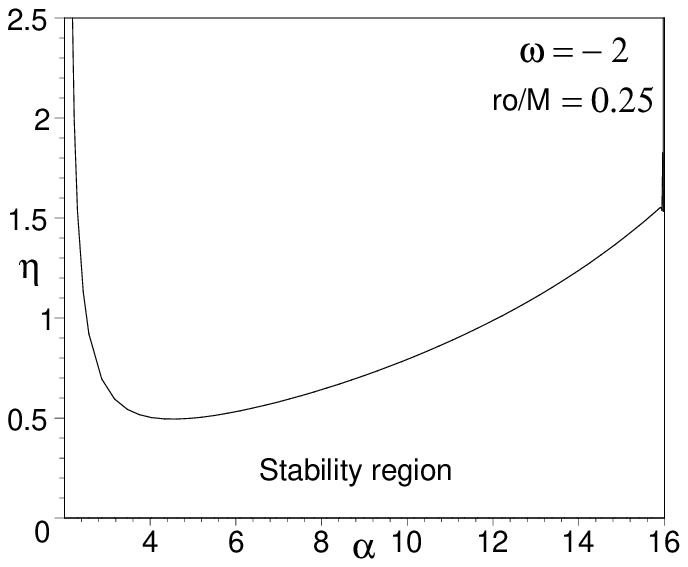}
  \caption{Plots for a positive surface energy density, i.e.,
  $b(a_0)<2M$. We have defined $\alpha=a_0/M$, and considered
  $\omega=-2$ for both cases. The first plot is given by $r_0/M=1$,
  and the second by $r_0/M=0.25$.
  The stability regions are given below the solid curve.
  See the text for details.}
  \label{phantomWH}
\end{figure}

For a fixed value of the parameter, for instance $\omega=-2$, we
shall consider the following cases: $r_0/M=1.0$, so that
$2<a_0/M<4$; and for $r_0/M=0.25$, we have $2<a_0/M<16$. The
respective stability regions are depicted in Fig. \ref{phantomWH}.
From Eq. (\ref{consequation2}), one may show that $\sigma'<0$, and
as $\sigma>0$, this implies $m_s \sigma'<0$. Thus, the stability
regions, dictated by the inequality (\ref{stability2}), lie
beneath the solid lines in the plots of Fig. \ref{phantomWH}. For
decreasing values of $r_0/M$, as $a_0$ increases, the values of
$\eta_0$ are further restricted. Using positive surface energy
densities, stable geometries may be found well within the bound of
$0<\eta_0 \leq 1$, and the stability regions increase for
increasing values of $r_0/M$.
\begin{figure}[t]
\centering
  \includegraphics[width=2.3in]{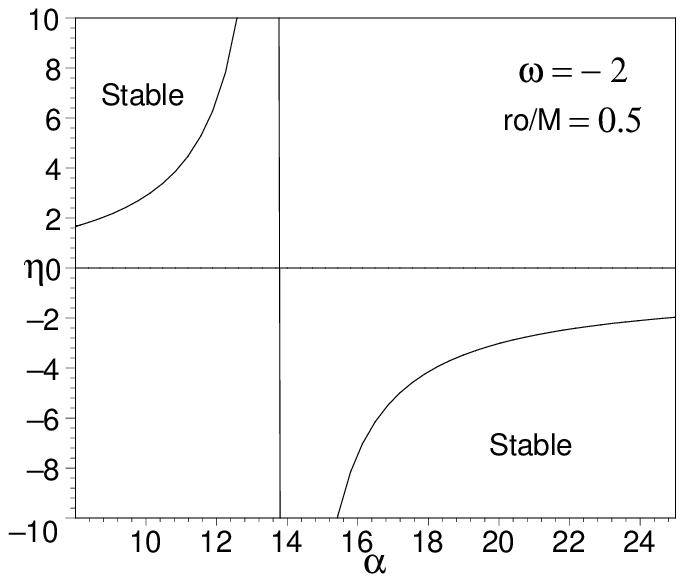}
  \hspace{0.4in}
  \includegraphics[width=2.3in]{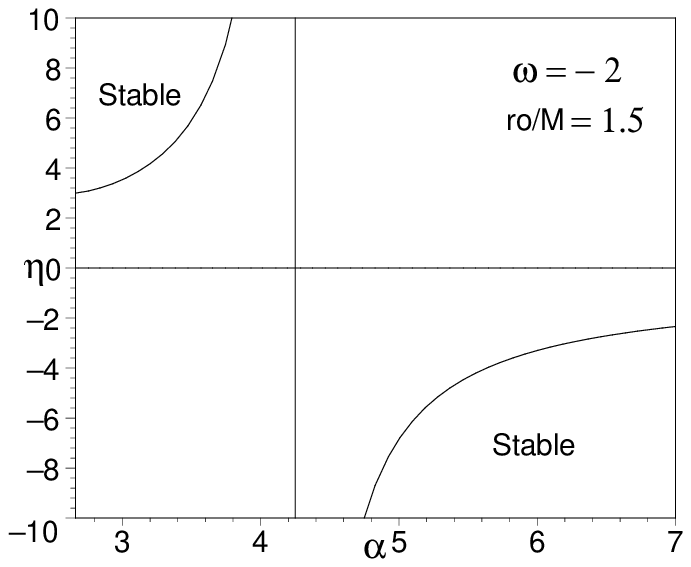}
  \caption{Plots for a negative surface energy density,
  considering $r_0/M<2$.
  We have defined $\alpha=a_0/M$, and considered
  $\omega=-2$ for both cases. The first plot is given by
  $r_0/M=0.5$, and the second by $r_0/M=1.5$.
  The stability regions are given above the first
  solid curve, and below the second solid curve.
  See the text for details.}
  \label{phantomWH2}
\end{figure}

For $b(a_0)>2M$, the surface mass of the thin shell is negative,
$m_s(a_0)<0$. We shall separate the cases of $r_0<2M$ and
$r_0>2M$.
If $r_0<2M$, the range of the junction radius is given by $a_0 >2M
\left(2M/r_0\right)^{-(1+\omega)}$. For this specific case,
$\sigma'$ possesses one real positive root, $R$, in the above
range, signalling the presence of an asymptote, $\sigma'|_R=0$. We
verify that $\sigma'<0$ for $2M(2M/r_0)^{-(1+\omega)}<a_0 <R$, and
$\sigma'>0$ for $a_0
>R$. Thus, the stability regions are given by
\begin{eqnarray}
\eta_0 &>& \overline{\Theta}, \quad {\rm if} \quad  2M
\left(2M/r_0\right)^{-(1+\omega)}<a_0 <R  \,,
         \label{flat-stability1}
       \\
\eta_0 &<& \overline{\Theta}, \quad {\rm if} \quad a_0 >R\,.
        \label{flat-stability2}
\end{eqnarray}

Consider for $\omega=-2$, the particular cases of $r_0/M=0.5$, so
that $a_0/M>8$, and $r_0/M=1.5$, so that $a_0/M>2.667$. The
asymptotes, $\sigma'|_R=0$, for these cases exist at $R/M \simeq
13.9$ and $R/M\simeq 4.24$, respectively. These cases are
represented in Fig. \ref{phantomWH2}. Note that for increasing
values of $r_0/M$, the range of $a_0$ decreases, and the values of
$\eta_0$ are less restricted. Thus, one may conclude that the
stability regions increase, for increasing values of $r_0/M$.

If $r_0>2M$, then obviously $a_0>r_0$. We verify that $\sigma'>0$,
and consequently $m_s\,\sigma'<0$, so that the stability region is
given by inequality (\ref{stability2}). We verify that the values
of $\eta_0$ are always negative. However, by increasing $r_0/M$,
the values of $\eta_0$ become less restricted, and the range of
$a_0$ decreases.

\subsubsection{Isotropic pressure, $p_r=p_t=p$}

Consider an isotropic pressure phantom wormhole geometry, with the
redshift function given by Eq. (\ref{isotropic-redshift}),
$\Phi(r)=(3\omega+1)/(1+\omega)\;\ln \left(r/r_0\right)$ and
$b(r)=r_0\,(r/r_0)^{-1/\omega}$.
To determine the stability regions of this solution, as in the
previous case, we shall separate the cases of $b(a_0)<2M$ and
$b(a_0)>2M$.

For $b(a_0)<2M$, we have $m_s >0$, and the condition $r_0<2M$ is
imposed. Therefore, the junction radius lies in the same range as
the previous case, i.e., Eq. (\ref{flat-range}). We also verify
that $\sigma' <0$ in the respective range. Thus the stability
region is given by
\begin{equation}
\eta_0 < \overline{\Theta}, \quad {\rm if} \quad 2M< a_0 <2M
\left(2M/r_0\right)^{-(1+\omega)}  \,.
\end{equation}

\begin{figure}[t]
\centering
  \includegraphics[width=2.3in]{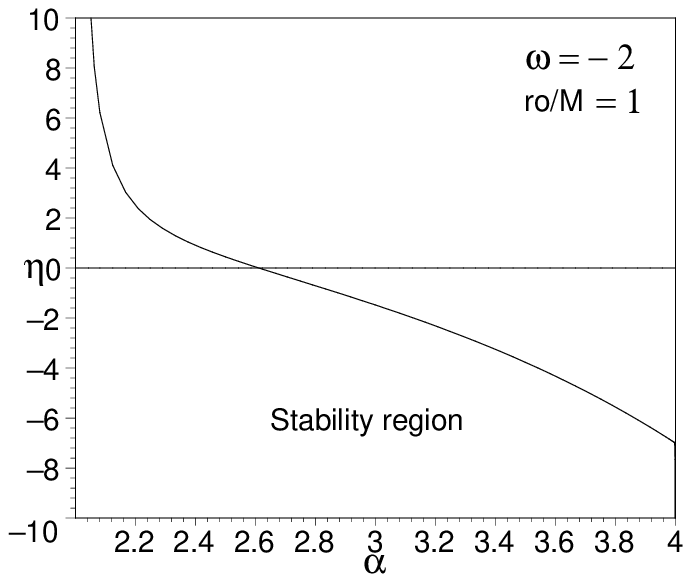}
  \hspace{0.4in}
  \includegraphics[width=2.3in]{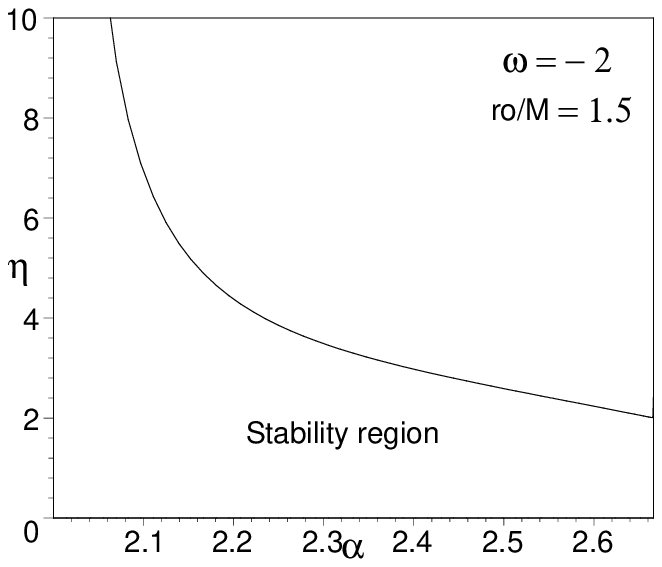}
  \caption{Plots for an isotropic pressure phantom wormhole.
  We have defined $\alpha=a_0/M$ and considered $\omega=-2$
  for both cases. For $b(a_0)<2M$, the condition $r_0<2M$ is imposed.
  The first plot is given by $r_0/M=1.0$, and the second by $r_0/M=1.5$.
  The stability regions are given below the solid curves.}
  \label{iso1}
\end{figure}

Consider, for simplicity, $\omega=-2$, and the cases for $r_0/M=1$
and $r_0/M=1.5$ are analyzed in Fig. \ref{iso1}. The ranges are
given by $2<a_0/M<4$ and $2<a_0/M<2.667$, respectively. Note that
as $r_0/M$ decreases, the range of $a_0$ increases. However, the
values of the parameter $\eta_0$ become more restricted. Thus, one
may conclude that the stability regions increase, as $r_0/M$
increases.

For $b(a_0)>2M$, then $m_s(a_0)<0$. As before, we shall separate
the cases of $r_0<2M$ and $r_0>2M$. For $r_0<2M$, the range of
$a_0$ is given by $a_0 >2M (2M/r_0)^{-(1+\omega)}$, as in the
previous case of the asymptotically flat wormhole spacetime.

\begin{figure}[t]
\centering
  \includegraphics[width=2.3in]{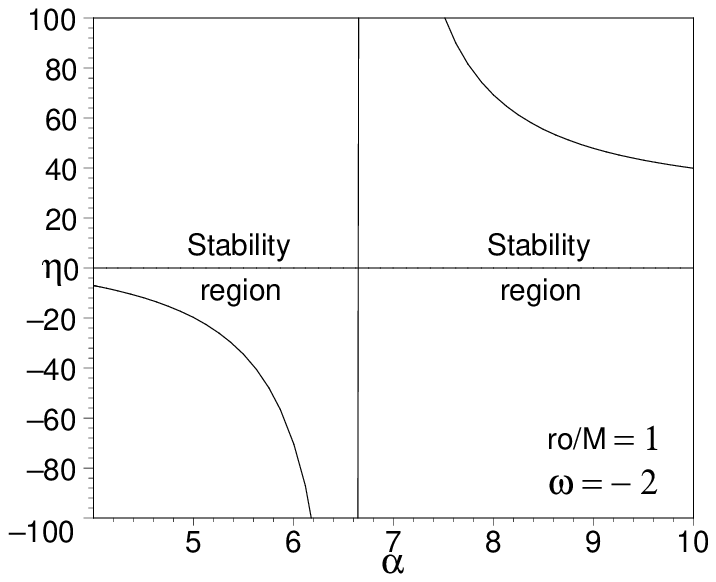}
  \hspace{0.4in}
  \includegraphics[width=2.3in]{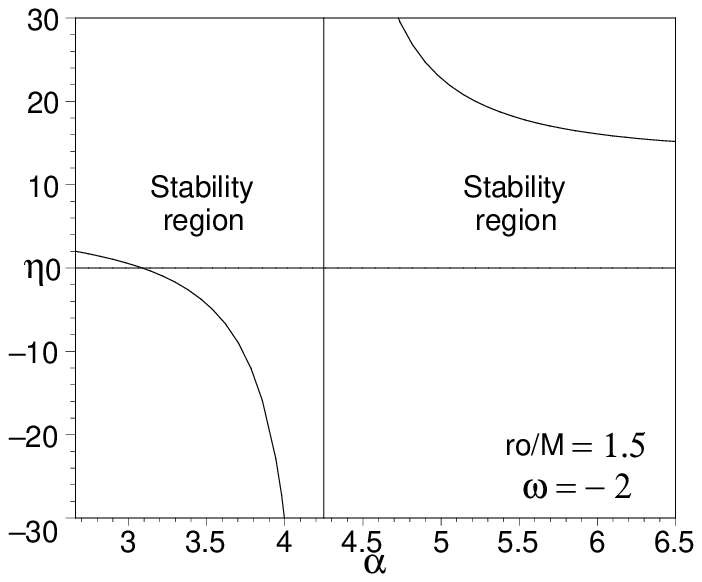}
  \caption{Plots for an isotropic pressure phantom wormhole,
  for $b(a_0)>2M$ and $r_0<2M$. We have defined $\alpha=a_0/M$ and
  considered $\omega=-2$ for both cases. The first plot is given
  by $r_0/M=1.0$, and the second by $r_0/M=1.5$.
  The stability regions are given above the first solid curve,
  and below the second solid curve.}
  \label{iso2}
\end{figure}

For this case $\sigma'$ also possesses one real positive root,
$R$, in the respective range. We have $\sigma'<0$ for
$(2M/r_0)^{-1/\omega}<a<R$, and $\sigma'>0$ for $a_0>R$. The
stability regions are also given by the conditions
(\ref{flat-stability1})-(\ref{flat-stability2}). We have
considered the specific cases of $r_0/M=1$ so that the respective
range is $a_0/M>4$; and $r_0/M=1.5$, so that $a_0/M>2.667$. The
asymptotes, $\sigma'|_R=0$, for these cases exist at $R/M\simeq
6.72$ and $R/M \simeq 4.24$, respectively. This analysis is
depicted in the plots of Fig. \ref{iso2}. For decreasing values of
$r_0/M$, the stability parameter $\eta_0$ becomes less restricted
and the range of the junction radius increases. Thus, one may
conclude that the stability regions increase for decreasing values
of $r_0/M$.

\section{Conclusion}\label{Sec:conclusion}

As the Universe is probably constituted of approximately 70\% of
null energy condition violating phantom energy, this cosmic fluid
may be used as a possible source to theoretically construct
traversable wormholes. In fact, it was found that infinitesimal
amounts of phantom energy may support traversable wormholes
\cite{Lobo-phantom}. In this paper, we have modelled phantom
wormholes by matching an interior traversable wormhole geometry,
satisfying the equation of state $p=\omega \rho$ with $\omega<-1$,
to an exterior vacuum solution at a finite junction interface. We
have analyzed the stability of these phantom wormholes, an issue
of fundamental importance, to linearized perturbations around
static solutions, by including the momentum flux term in the
conservation identity. We have considered two particularly
interesting cases, namely, that of an asymptotically flat
spacetime, and that of an isotropic pressure wormhole geometry. We
have separated the cases of positive and negative surface energy
densities and found that the stable equilibrium regions may be
significantly increased by strategically varying the wormhole
throat. As we are considering exotic matter, we have relaxed the
condition $0 < \eta_0 \leq 1$, and found stability regions for
phantom wormholes well beyond this range.

An interesting feature of the phantom regime is that due to the
fact of the accelerated expansion of the Universe, macroscopic
wormholes could naturally be grown from the submicroscopic
constructions that originally pervaded the quantum foam
\cite{gonzalez2}.
The accretion of phantom energy onto Morris-Thorne wormholes was
further explored in Refs.~\cite{diaz-phantom3}, and it was shown
that this accretion gradually increases the wormhole throat which
eventually overtakes the accelerated expansion of the universe,
consequently engulfing the entire Universe, and becomes infinite
at a time in the future before the big rip. This process was
dubbed the ``Big Trip'' \cite{diaz-phantom3}.
Traversable wormholes supported by the generalized Chaplygin gas
were also found \cite{ChapWH}, and in the context of the ``Big
Trip'', it was found that the latter may be avoided in a wide
region of the Chaplygin parameters, by the accretion of a
generalized Chaplygin gas onto wormholes \cite{Madrid2}.
In concluding, it is noteworthy the relative ease with which one
may theoretically construct traversable wormholes with the exotic
fluid equations of state used in cosmology to explain the present
accelerated expansion of the Universe.



\end{document}